# The Levels of Conceptual Interoperability Model: Applying Systems Engineering Principles to M&S

**Wenguang WANG**

**National University of Defense Technology, Changsha 410073, China**

wgwangnudt@gmail.com

**Andreas TOLK[+]**

**Old Dominion University, Norfolk, VA 23529, United States**

atolk@odu.edu

**Weiping WANG**

**National University of Defense Technology, Changsha 410073, China**

wangwp@nudt.edu.cn

**Abstract**

This paper describes the use of the Levels of Conceptual Interoperability Model (LCIM) as a framework for conceptual modeling and its descriptive and prescriptive uses. LCIM is applied to show its potential and shortcomings in the current simulation interoperability approaches, in particular the High Level Architecture (HLA) and Base Object Models (BOM). It emphasizes the need to apply rigorous engineering methods and principles and replace ad-hoc approaches.

## 1. INTRODUCTION

Modeling and Simulation (M&S) is an emerging discipline. While modeling comprises the conceptual work of defining the model by abstracting its data, processes, and constraints from reality, simulation focuses on the implementation of such models as executables, usually in the form of computer programs. As with any emerging discipline, the initial work is driven by the accomplishments of individuals. However, in order to make M&S a scientific discipline, a body of knowledge needs to be established that comprises engineering methods in support of standard operations. Similar to system architecture and modeling concepts, as they have been established by Systems Engineering, frameworks and methods are necessary to capture and document our simulation systems. As the operational concept is part of the system description, the conceptualization must also be part of the simulation documentation. In order to support identification, selection, and composition of models and simulation systems, current practices are not sufficient.

Within this paper, the Levels of Conceptual Interoperability Model (LCIM) was evaluated to see to what degree it can support the different necessary artifacts. The results are based on research that tried to compose the different aspects highlighted in various publications on the LCIM into an overview and to derive new insights. It is first used as a frame for conceptual modeling. Second, its applicability as a descriptive as well as a prescriptive model is evaluated. Finally, it is applied to look at current simulation interoperability standards, in particular the High Level Architecture (HLA) and Base Object Models (BOM).

## 2. INTRODUCING A SYSTEMS ENGINEERING FRAMEWORK FOR SIMULATION INTEROPERATION

Systems Engineering defines several principles that support the design of systems and documenting their functionality and interfaces in a way ensuring that independently designed systems can interoperate with each other. Within this section, we will introduce the ideas of conceptual modeling and capturing the resulting artifacts in a systematic way, using the LCIM as the guiding frame, in support of simulation interoperation.

### 2.1. Conceptual Modeling and Interoperability

A conceptual model is the abstract and simplified representation of systems for some specific purpose by languages, figures, tables, or other suitable artifacts. Robinson defines a conceptual model as *"a non-software specific description of the simulation model that is to be developed, describing the objectives, inputs, outputs, content, assumptions, and simplifications of the model."* [1] Conceptual modeling is the process of creating a conceptual model. Based on his research, Robinson proposes a definition, requirements, and conceptual modeling framework.[1] Following Robinson's proposal, conceptual models and conceptual modeling:

- reduce ambiguity, incompleteness, inconsistency, and mistakes in the description of requirements,
- facilitate the communication between stakeholders in modeling and simulation processes, such as users, architects, analyzers, domain experts, modelers, and developers,
- form the basis and start for successive phases (analysis, design, implementation and application),
- facilitate the Verification, Validation and Accreditation (VV&A) of models and simulation systems, and
- promote the reusability, interoperability, and composability of simulation resources.

As such, conceptual modeling becomes a key enabler for Modeling and Simulation (M&S) services/components that are interoperable and composable. Petty and Weisel define composability as *"the capability to select and assemble simulation components in various combinations into valid simulation systems to satisfy specific user requirements."* [2] Furthermore, they distinguish between syntactic and semantic composability[3]. Hofmann observes that to compose simulation systems meaningfully and achieve valid interoperability among the simulation systems and underlying models, the alignment and consistent comprehension should be reached at the conceptual model level[4]. Similar observations were made within the Simulation Interoperability Standards Organization (SISO) within their Conceptual Modeling Study Group[5].

To understand why conceptual models are important to interoperability and composability, it is necessary to understand how machines gain understanding: they need a consistent system description in form of machine-understandable meta-data regarding data, processes, and constraints[6]. The system – or the M&S service – is herein described by its properties that are grouped into propertied concepts (the basic simulated entities and attributes), the processes (the behavior of simulated entities and how their attributes change), and constraints (assumptions constraining the values of the attributes and the behavior of the system).

Similar ideas were first introduced by Zeigler when he published a model for understanding a system within another observing system[7]. He introduces three premises that need to be supported by the annotations describing the M&S services.

- The first premise is that the observing system has a *perception* of the system to be understood. This means that the properties and processes must be observable and perceivable by the observing system. The properties used for the perception should not significantly differ in scope and resolution from the properties exposed by the system under observation.
- The second premise is that the observing system needs to have a *meta-model* of the observed system. The meta-model is a description of properties, processes, and constraints of the expected behavior of the observed system. Without such a model of the system, understanding is not possible. In addition, it allows the internal "simulation of the observed system" to support machine understanding of what the observed system will do next (anticipation) or is capable of.
- The third premise is the *mapping* between observations resulting in the perception and meta-models explaining the observed properties, processes, and constraints.

In summary, to gain meaningful interoperability and composability, descriptions of the service are needed allowing for the creation of the perception of what - and how - the service will work out in support of understanding. This description must comprise the elements proposed by Robinson. In other words, the description must capture the objectives, inputs, outputs, content, assumptions, and simplifications of the models. This is where conceptual model works.

However, it is not easy to capture the elements of conceptual models. First, there are many definitions, contents, and understanding on conceptual modeling which have not been widely accepted and standardized. Furthermore, Robinson stated that there is no right conceptual model for any specified problem because the model is an agreement between more than one person (the modeler, clients and domain experts). Moreover, the three premises proposed by Zeigler are hard to reach even between two people (implicated intent and meaning beyond words can be hard to understand and process). Therefore the conceptual interoperability between two machines is hard to accomplish. Can conceptual interoperability and composability really be achieved?

Although it is a difficult problem, it should be attempted, even if it will never be reached. Every step closer allows for lessons learned and helps the repetition of avoidable mistakes. Although we may not be able to prove that two systems are conceptually interoperable, we may find ways to point to critical areas or to show when two systems are NOT conceptually interoperable. Additionally, we can use the idea of "divide and conquer"; go towards the conceptual interoperability goal step by step using hierarchical approaches. The LCIM[8] is the method and model towards conceptual interoperability and composability with these ideas chosen within this paper as a promising engineering method.

**2.2. Levels of Conceptual Interoperability Model**

Systems Engineering requires a reproducible and well documented approach. In other words: a framework is required to capture the artifacts needed for simulation interoperation. LCIM was proposed to deal with conceptual interoperability issues beyond technical interoperability. There are several levels of interoperability models in technical domain such as Levels of Information Systems Interoperability (LISI) model, the NATO Model for Interoperability etc.[9]. However, meaningful interoperability of simulation systems on the implementation level requires composability of the underlying conceptual models[8]. LCIM divides conceptual interoperability into layers to cope with these problems.

LCIM was originally proposed by Tolk and Muguira[8]. After continuous evolution, it forms the latest version illustrated in Figure 1. [10-12]

The seven levels from 'no interoperability' to 'conceptual interoperability' are notated from L0 to L6, whose implications are listed in Table 1. Bottom-up refers to from L0 to L6, top-down vice versa.

**Table 1.** Implications of LCIM

| Level | Layer Name | Premise | Information defined | Contents clearly defined | Domain | Focus | Capability |
|---|---|---|---|---|---|---|---|
| L6 | Conceptual | Common conceptual model | Assumptions, constrains etc. | Documented conceptual model | Modeling abstraction | Composability | High |
| L5 | Dynamic | Common execution model | Effect of data | Effect of information exchanged | | | |
| L4 | Pragmatic | Common workflow model | Use of data | Context of information exchanged | Simulation implementation | Interoperability | Medium |
| L3 | Semantic | Common reference model | Meaning of data | Content of information exchanged | | | |
| L2 | Syntactic | Common data structure | Structured data | Format of information exchanged | | | |
| L1 | Technical | Common communication protocol | Bits and bytes | Symbols of information exchanged | Network connectivity | Integratability | Low |
| L0 | No | No connection | NA | NA | | | |

After several years of development, LCIM is becoming more and more mature and is gaining more recognition. *(In November 2008, LCIM was referenced in Wikipedia[13], and LCIM has been cited 79 times according to Google scholar's statistics[14].)* Besides the simulation interoperability community, LCIM is used by scientists of multiple disciplines to deal with problems in their communities[10], e.g. system biologist and ontology researchers. This shows that LCIM has wide application potential.

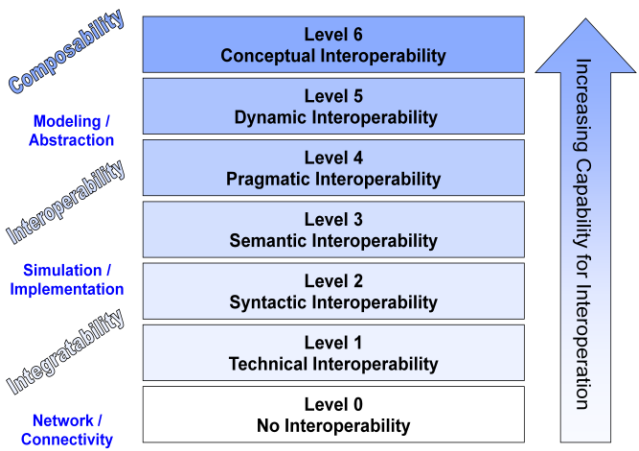

**Figure 1.** The Levels of Conceptual Interoperability Model

## 3. DESCRIPTIVE AND PRESCRIPTIVE ROLES

Systems engineering distinguishes between explaining (descriptive) and mandating (prescriptive) models. LCIM can serve in both functions[11]: In its descriptive role, LCIM describes the levels and properties of interoperability existing within a composition of systems. In the prescriptive role, LCIM prescribes the methods and requirements that must be satisfied during the engineering of a system of systems to achieve a desired level of interoperability. Referring to the definitions of descriptive linguistics[15] and prescriptive linguistics[16] in addition to the definitions of descriptive and prescriptive models by Department of Defense (DoD)[17], we define the descriptive and prescriptive roles of LCIM and clarify their relationships as the basis for later investigation.

### 3.1. Descriptive Role

Referring to the definition of descriptive model by the DoD[17], we define the descriptive role of LCIM as the role used to depict or analyze the ability, properties, characteristics and the levels of conceptual interoperability of an existing system or system of systems. In this case, LCIM serves as a documentation and maturity model. The goal of descriptive role is to describe how existing systems are interoperating and what level of conceptual interoperability can be reached by user's specific approaches without prescription. The outcome of descriptive role can be used to evaluate the interoperability of existing systems and inform the users of the current properties and capabilities of interoperability. The following cases can be identified by the descriptive role: It is possible that two models are derived from exactly the same concepts, but their implementations differ. It is also possible, that the same entities are used in the simulation, but represent different concepts. It also shows technical problems, like one system using UDP/IP, the other using TCP/IP. In the descriptive role, LCIM is used to show the gaps that need to be closed.

The characteristics of the descriptive role are listed as:
- Real systems or system of systems that have been implemented or existed;
- The specific technical approaches, implementations and documentations are known;

- LCIM is used as a maturity model. When higher levels are reached, the lower levels must have been satisfied;
- The levels are mutual supportive bottom-up. The lower level is the premise of the higher and the higher needs the implementation of lower levels.

The contents of descriptive role at each level are shown in Table 2[11].

**Table 2**. Descriptive Role of LCIM

| Levels | Description of Interoperability at this level |
| --- | --- |
| L6(Conceptual) | Interoperating systems at this level are completely aware of each others information, processes, contexts, and modeling assumptions. |
| L5(Dynamic) | Interoperating systems are able to re-orient information production and consumption based on understood changes to meaning, due to changing context as time increases. |
| L4(Pragmatic) | Interoperating systems will be aware of the context (system states and processes) and meaning of information being exchanged. |
| L3(Semantic) | Interoperating systems are exchanging a set of terms that they can semantically parse. |
| L2(Syntactic) | Have an agreed protocol to exchange the right forms of data in the right order, but the meaning of data elements is not established. |
| L1(Technical) | Have technical connection(s) and can exchange data between systems |
| L0(No) | NA |

The use and processes of descriptive role are describe the documentation, specific approach, and implementation of interoperating systems from L0 to L6 according to Table 2, and evaluate the levels they can reach. If the recommended approaches in Table 3 are used in the implementation, then the systems can be described and judged by referring to the prescription role in Section 3.2.

### 3.2. Prescriptive Role

Referring to the definition of prescriptive model by DoD[17], we define the prescriptive role of LCIM as the role used to depict and prescribe the ability, properties, characteristics and the levels of conceptual interoperability of a proposed system or system of systems. In other words, it prescribes the approaches and requirements that must be satisfied to accommodate a target degree of conceptual representation between systems. In this case, LCIM serves as a conceptual interoperability guidance model. The interoperability or real systems may even not exist. As a prescription, LCIM gives users a metric of requirements to reach a certain level of conceptual interoperability.

The characteristics of the prescriptive role are listed as the following:
- Real systems or system of systems do not exist or in the phase of conceptual modeling;
- As an interoperability guidance model, LCIM can be used to recommend the required conditions and possible engineering approaches to reach a certain interoperability level;
- Top-down mappings are needed from conception to implementation. If problems are only dealt with at the conceptual level without any real systems, the lower levels are unreachable until real systems are implemented;
- The levels are mutually supportive bottom-up, mapped, and refined top-down. Higher levels need the implementation of the lower. There are much work to do for the mapping from the higher to the lower;
- Only reaching the higher level is not enough, full spectrum interoperability at all the levels is needed especially for the prescriptive role of LCIM;
- As the development of techniques, the recommended approaches by the prescriptive role are ever changing and evolving and only have relative stability in a certain period.

In practice, systems are often developed from conceptual models to design models and final implementation models and applications or define objectives, perform conceptual analysis, design federation and develop federation process in FEDEP. The prescriptive role of LCIM is used to facilitate the transformation from conceptual modeling to systems implementation focusing on the interoperation aspects. LCIM is a guide - or check list - how to reach the target. In this case, it builds logically top-down.

Referring to the literature[11,18,19], the contents of prescriptive role at each level are illustrated in Table 3. From which Ontology, Unified Modeling Language (UML), Model Driven Architecture (MDA), and Discrete Event System Specification (DEVS) are the important approaches to enhance the conceptual interoperability and composability.

The use and processes of prescriptive role are synthesizing and balancing the goal, scheduling, conditions, costs etc., deciding the highest level that should be reached.; determining the information exchange requirements and choosing the proper approaches and data structures, semantics and so on according to Table 3; refining problems and objectives, analyzing and mapping top-down and select approaches at lower levels to guide the design and implementation.

### 3.3. Comparison and Relationship between Descriptive and Prescriptive Roles

The descriptive and prescriptive roles apply to LCIM through different viewpoints. To reach semantically lossless interoperation, all levels are needed in both cases. Description and prescription have differences and similarities as illustrated in Table 4.

**Table 3**. Prescriptive Role of LCIM

| Levels | Prescription of Requirements to reach this Level | Common Reference Engineering Approaches |
| --- | --- | --- |
| L6(Conceptual) | A shared understanding of the conceptual model of a system (exposing its information, processes, states, and operations). | DoDAF; Military Mission to Means Framework; Platform Independent Models of the Model Driven Architecture; SysML |
| L5(Dynamic) | The means of producing and consuming the definitions of meaning and context is required. | Ontology for Services; UML artifacts; DEVS; complete UML; BOM |
| L4(Pragmatic) | A method for sharing meaning of terms and methods for anticipating context are required. | Taxonomies; Ontology; UML artifacts, in particular sequence diagrams; DEVS; OWL; MDA |
| L3(Semantic) | Agreement between all systems on a set of terms that grammatically satisfies the syntactic level solution requirements is required. | Common Reference Model, such as C2IEDM and CADM; Dictionaries; Glossaries; Protocol Data Units; RPR FOM |
| L2(Syntactic) | An agreed-to protocol that all can be supported by the technical level solution is required. | XML; HLA OMT; Interface Description Language; CORBA; SOAP |
| L1(Technical) | Ability to produce and consume data in exchange with systems external to self is required. | Network connection standards such as HTTP; TCP/IP; UDP/IP etc. |
| L0(No) | NA | NA |

**Table 4**. Comparisons between the Descriptive and Prescriptive Roles of LCIM

| Items | Descriptive Role | Prescriptive Role | Relationship |
| --- | --- | --- | --- |
| Purpose | Act as interoperability maturity model to evaluate the interoperability and composability of existing systems | Act as interoperability guidance model to prescribe and guide the interoperability and composability design and implementation of proposed systems | 1. Both base on the basic levels and meanings of LCIM |
| Form of objects | Existing systems may have interoperability | Proposed systems without interoperability | 2. Both concentrate on and are limited to the data exchange focus |
| Stakeholder | Users, VV&A users | Decision makers and designers | 3. Both have some similar content. |
| Applicable phase | Late phases such as VV&A and after action review etc. | Early phases such as requirements analysis and conceptual design etc. | 4. Complementarily: Prescription and description are essentially complementary and can be combined for certain purposes. |
| Utility | Evaluate the degree of conceptual representation between interoperating systems. Accept and evaluate specific approaches actually used. | Provide a metric of requirements to reach a certain degree of conceptual interoperability. Recommend the required conditions and possible engineering approaches. | 5. Conflict: Collisions may occur when reality does not conform to expectation, especially when integrating existing systems to system of systems for a certain conceptual interoperability goal. Compelling means may be used by government for these circumstances. |
| Principle | Describe what data and how current systems exchange in practice. | Guide the proposed systems as to how and what data and form should be exchanged. | |
| Content | Objectively describe current interoperability without comments or bias. Accept non-standard and specific approaches. | Prescribe the requirements, rules, recommendations, or avoidable mistakes for systems to reach a certain level. Try to avoid non-standard approaches. Improvements to old standard approaches can be accepted when they are standardized. | |
| Application Context | Various applications used by private community or individuals. Not constrained by government standards or formal circumstances. | Education, publishing or compulsory formal standards or circumnutates by government sponsors (e.g. The compelling use of HLA standard by DoD) | 6. When specific methods described by the descriptive role prevail and are standardized, they may be accepted by the prescriptive role. |
| Process | Check and aggregate bottom-up | Refine and implement top-down | |

The combinations of the two roles can facilitate better solutions to interoperation and composition problems. There are two kinds of combinations. One is from scratch without any existing and implemented systems. Prescription can be first used with the analysis to the interoperation objectives, costs, conditions and so on. As the refinement and implementation gradually, the descriptive role can be used to evaluate the interoperability of that process. The gaps from reality to objective must be found to facilitate the design and implementation in later processes. The prescription - description - prescription again - description again iterative cycle can be used in this case. The other kind of combination is reuse-based. Some systems or subsystems have existed or been implemented and need to be integrated to reach a certain interoperation goal. The description and evaluation - prescription - objective execution - description - evaluation iterative cycle can be used. In both cases, LCIM can be seen as the two sides of a coin - one is what exist or have done, the other is what needs to be done.

## 4. APPLICATION OF LCIM TOWARD THE COMPOSABLITY OF HLA AND BOM

In this section, HLA and BOM are taken as the examples to illustrate the use of descriptive and prescriptive roles of LCIM to evaluate, analyze, and recommend composable solutions formulated in the respective paradigms. The descriptive role is used to analyze composability of existing systems using HLA or BOM. Next, the relationship between LCIM and BOM is presented. This shows that BOM is the important step toward conceptual interoperability and composability. Finally, the prescriptive role is used to give some recommendations to improve the composability of BOM and HLA.

### 4.1. Current Composability of HLA and BOM

HLA[20] is an IEEE simulation standard and widely used as a common simulation framework to support the interoperability and reusability of various simulation applications. BOM[21] is a component-based simulation object specification to improve the composability, reusability, and interoperability at the conceptual model level. BOM can be used within the HLA approach, but BOM is independent from HLA and has been successfully applied using other approaches as well.

Driven by the extension of the application scope, the development of new technology and the need of net-centric simulation in Global Information Grid, many deficiencies on interoperability, extensibility, and reusability of HLA have been revealed during the past decade. Meanwhile, to facilitate the reusability of simulation resources and rapid development of applications, composable simulation[3,8,22,23] has drawn significant attention in M&S community in recent years. As the leading standard in the distributed simulation community, HLA should also be extended to improve composability itself. A recent peer survey[24] reveals that the practical relevance and revision of HLA are still the future trends in distributed simulation.

The composability of HLA focuses on the object models. BOM facilitates the composability of HLA framework at the conceptual level in general. Other approaches, such as the modular Federation Object Model / Simulation Object Model (FOM/SOM)[25,26] approach, are still focusing on the implementation level. While a strong connection to HLA is not a negative trend *per se*, alternative domains, such as service-oriented architectures and Web Services, will be of increasing importance in the future [27,28].

With the descriptive role of LCIM, the level of interoperation for BOM and HLA solutions, such as FOM/SOM, FOM/SOM module, Management Object Model (MOM), Real-time Platform Reference (RPR)-FOM, and RPR-BOM, can be evaluated in Figure 2. The vertical axis stands for the increasing levels of interoperation, while the horizontal axis represents the increasing component granularity of object models.

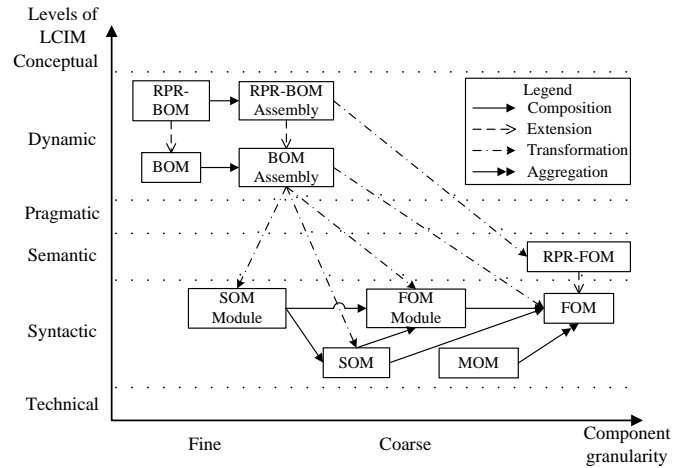

**Figure 2**. Composability of BOM and HLA Object Models

There are four kinds of relationships depicted in Figure 2.
- Composition relationship means that several fine granular components can be composed to a course one.
- Extension relationship stands for some extensions that can be made on standard object models to fit specific applications.
- Transformation relationship refers to BOM assembly, which can be transformed to FOM/SOM modules or FOM/SOM by mechanisms such as XSLT.
- Aggregation relationship refers to MOM - each FOM has one and only one.

According to the descriptive role of LCIM, the Runtime Infrastructure (RTI) implementations used by HLA federations systems generally adopt TCP/IP, UDP/IP, or HTTP protocols, which are at the technical level. FOM is the core of HLA federations. FOM/SOM modules, SOM and MOM can be composed to FOM. They all conform to the HLA OMT specification. OMT specifies the structure but not semantic of data exchanged, hence FOM/SOM, FOM/SOM module and MOM can only reach the syntactic level. RPR-FOM specifies standard common information exchange reference models. Users can gain a common agreement allowing RPR-FOM to reach the semantic level. BOM targets the conceptual level, but the assumptions and constrains can not be easily captured by the UML artifacts that BOM uses. Hence BOM can not reach the conceptual level. However, the UML models used allow it at least to get up to the dynamic level.

The level of interoperation supported by HLA can be concluded from Table 2. At present, HLA can only reach the syntactic level in general. Especially FOM, the core of HLA, constrains the interoperability and composability. HLA is good at simulation interoperability, but lays little emphasis on model composability. It can further be seen from Figure 2 that modular FOM/SOM refines the composable granularity and improves the composability. The use of RPR-FOM increases the HLA level of interoperation up to the semantic level, but only within the application range of this specific FOM. BOM refines the composable granularity and greatly improves the levels of interoperation, as the standard requires the rigorous definition of artifacts on all levels, including first elements for the conceptual level.

### 4.2. Improving BOM using LCIM

By integrating artifacts of conceptual models and implementation details into the standard, BOM improves composability, interoperability, and reusability. As such, BOM plays an important role to improve the composability of HLA without being limited to the HLA domain. Meanwhile, LCIM is a good theory or model to evaluate the levels of interoperability and composability. The relationships between BOM and LCIM need further research.

The BOM standard uses artifacts to describe the supported M&S service on all levels. These artifacts describe entities, events, and aspects of behavior, which was defined by patterns of interplay and state machines. In addition, the object models interfaces, model mappings from conceptual models to object models and metadata are also included.

Research on BOM conducted at the authors' institutes show that the focus BOM lays on the conceptual, semantic, syntactic levels and the mappings among them. XML expression makes BOM at least up to the syntactic level. Moradi et.al. proposed a method using OWL to enrich BOM semantics[29]. Mojtahed et. al. presented two methods to semantically extend BOM[30]. There are many compositions to use the four parts of BOM (metadata, conceptual model, mapping conversation and object class interface)[31]. They are all dedicated to making BOM cover wider and higher conceptual interoperability levels.

In summary, BOM can be seen as a bridge between the conceptual space to implementation space[32]. BOM comprises artifacts to address all levels of interoperation addressed by the LCIM. However according to the prescriptive role of LCIM and the elements defined by Robinson, BOM does not capture the assumptions and constrains of conceptual models, so it is not sufficient for composability as expressed by the conceptual level of the LCIM. It is not the fault of BOM but the limitation of UML. Hence BOM can only satisfy up to the dynamic level. However, BOM is the only simulation standard known to the authors that is not built exclusively on the data exchange paradigm but recognizes the fact that how and where the data are needed is important as well. Although the conceptual model of BOM needs to be improved, it is the most advanced simulation standard available.

Using ideas to capture data, processes, and constraints regarding scope, resolution, and structure on all levels of interoperation, as recently recommended by Tolk et al.[6], the BOM standard may be improved to become not only a system engineering frame for simulation, but also serves as a template providing the artifacts needed to support all levels of interoperation.

Using the prescriptive role of LCIM, even stronger suggestions to improve the levels of interoperation supported by HLA in general and by BOM in particular can be recommended. While description looks at each layer individually, prescription must ensure that all layers of interoperation up to the required level of interoperation are satisfied. BOM supports a structure to satisfy this requirement, but currently lacks some artifacts. The most obvious one is the inability to effectively cope with assumptions and constraints. Additional research regarding conceptual, pragmatic, and semantic artifacts of BOM supporting the idea of machine understandability as introduced by Zeigler is also required. Another topic requiring additional research is the possible combination of BOM and modular FOM/SOM[26] ideas, resulting in modular and reconfigurable BOM comprising all artifacts needed to support intelligent agents with the necessary metadata needed for identification, selection, composition, and orchestration of M&S services.

All these contributions also support HLA improvements. However, as previously stated, simulation interoperation is not limited to this domain. Alternative domains must remain a topic of research. Besides the approaches surveyed by Weisel[3], the MDA[33,34], the DEVS formalism with all its extensions[35,36], Ontology Research[11,35], and service-oriented architectures[34,36] are promising candidates to improving composability. Currently, HLA Evolved[37] targets are used to enhance and improve the interoperability, composability and reusability of HLA.

## 5. CONCLUSION

The LCIM started as a model to explain the needs to be specified to insure interoperation on different levels. These levels are all mutually supportive and form a sort of spectrum starting with the question "What do I model?" on the conceptual side and ending with the question "How do I model it?" on the technical side. Interoperation is only possible - in a meaningful way - if the technical structures are aligned with the conceptual ideas. As such, the LCIM is a framework describing a spectrum from what is conceptually modeled to how this is technically implemented. It requested that data (entities, propertied concepts) and how they are used (process concepts, methods) and the constraints are described by artifacts following the ideas of system architecture and modeling as known from the domain of system engineering. These descriptions must support the semantic loss-free mediation of viewpoints, as models may differ in scope, resolution, and structure on all levels.

The BOM standard is the first simulation interoperability standard that is not limited to pure data exchange specifications. While the artifacts used by BOM are insufficient to support all levels of interoperation as specified by LCIM, they build a solid foundation and the enhancement of BOM using the LCIM recommendations looks promising.

Evaluating the current research as reflected in journals and conference papers, there isn't much research and applications focusing on pragmatic and dynamic levels. Most research seems to be concentrating either on the conceptual level or on the semantic and syntactic level. While these papers address key issues in the processes of conceptualization as well as processes for design and implementation, they do not support a common understanding of states, processes, context etc. The framework proposed in this paper may contribute to merging the results in support of a common simulation engineering approach.

At first glance, the necessity for transparency required for composable services seems to conflict with the ideas of modularization and encapsulation. The loose coupling of services seems to contradict the rigorous requirements derived from the LCIM. However, if such artifacts are used, the concepts of loose coupling can still be supported while at the same time the composition of services that are conceptually not interoperable can be avoided. If this was not the case, technically feasible compositions are conceptually meaningless, and the results of such compositions are worthless. However, more research is needed in this domain as well.

The authors want to set a research agenda with this paper. Discussions and improvements are highly encouraged, as this can only be the beginning.

**Acknowledgement**

This work is partly supported by the National Natural Science Foundation of China (under Grant Nos 60674069, 60574056). The authors thank Paul Gustavson, Saikou Y. Diallo, Charles D. Turnitsa, Yifan Zhu and Qun Li for the valuable discussions.

**REFERENCES**

[1] Robinson S. "Conceptual modelling for simulation part I: Definition and requirements". *Journal of the Operational Research Society*, 2008, 59:278-290.
[2] Petty MD, Weisel EW. "A composability lexicon". *Spring Simulation Interoperability Workshop. Simulation Interoperability Standards Organization,* 2003.
[3] Weisel EW. "Models, composability, and validity" Thesis. *United States -- Virginia: Old Dominion University*, 2004.
[4] Hofmann MA. "Challenges of model interoperation in military simulations". *Simulation*, 2004, 80(12):659-667.
[5] Borah JJ. "SISO-REF-017-2006 simulation conceptual modeling (SCM) SG final report". 2006, http://www.sisostds.org
[6] Tolk A., Diallo SY, King RD, Turnitsa CD. "A Layered Approach to Composition and Interoperation in Complex Systems," Springer SCI 168, 2009, 41–74.
[7] Zeigler B. "Toward a simulation methodology for variable structure modeling". *In Elzas, Oren, Zeigler (Eds.) Modeling and Simulation Methodology in the Artificial Intelligence Era, North Holland.*, 1986.
[8] Tolk A, Muguira JA. "The levels of conceptual interoperability model". *Fall Simulation Interoperability Workshop. Orlando, Florida: Simulation Interoperability Standards Organization,* 2003.
[9] Morris E, Levine L, Meyers C, Place P, Plakosh D. "System of systems interoperability (SOSI), final report". *Software Engineering Institute, Carnegie Mellon University: P Pittsburgh*, 2004.
[10] Tolk A. "What comes after the semantic Web - PADS implications for the dynamic Web". *Proceedings of the 20th Workshop on Principles of Advanced and Distributed Simulation (PADS'06).* 2006.
[11] Tolk A, Turnitsa C, Diallo S. "Implied ontological representation within the levels of conceptual interoperability model". *Intelligent Decision Technologies*, 2008, 2:3-19.
[12] Tolk A, Diallo SY, Turnitsa CD. "Mathematical models towards self-organizing formal federation languages based on conceptual models of information exchange capabilities". *Winter Simulation Conference.* 2008.
[13] Wikipedia. "Conceptual interoperability". 2008/2008-06-19. http://en.wikipedia.org/wiki/Conceptual_interoperability.
[14] Google. "Google scholar". 2008/2008-11-9. http://scholar.google.com.
[15] Wikipedia. "Descriptive linguistics". 2008/2008-07-29. http://en.wikipedia.org/wiki/Descriptive_linguistics.
[16] Wikipedia. "Linguistic prescription". 2008/2008-07-29. http://en.wikipedia.org/wiki/Linguistic_prescription.
[17] "DoD modeling and simulation (M&S) glossary". *DoD 5000.59-M. United States: Defense Modeling and Simulation Office*, 1998. https://www.dmso.mil.

[18] Tolk A, Turnitsa CD, Diallo SY, Winters LS. "Composable M&S Web services for net-centric applications". *JDMS*, 2006, 3(1):27-44.
[19] Tolk A, Blais CL. "Taxonomies, ontologies, and battle management languages – recommendations for the coalition BML study group". *Spring Simulation Interoperability Workshop. Simulation Interoperability Standards Organization,* 2005.
[20] IEEE Std 1516.2-2000. "IEEE standard for modeling and simulation (M&S) high level architecture (HLA)-object model template (OMT) specification". *Standard*.
[21] SISO-STD-003-2006. "Base object model (BOM) template specification". *Standard*.
[22] Davis PK, Anderson RH. "Improving the composability of department of defense models and simulations". *Santa Monica: RAND Institute*, 2002.
[23] Ishom W. "Pursuit of composability and a direction towards a general framework to show composability". *Fall Simulation Interoperability Workshop.* 2007.
[24] Straßburger S, Schulze T, Fujimoto R. "Peer study final report: Future trends in distributed simulation and distributed virtual environments". 2008. http://www.sisostds.org/index.php?tg=fileman&idx=get&id=5&gr=Y&path=Standards+Activity+Committee%2FDocuments+for+Review&file=PeerStudy_DS_DVE_v1.0.pdf.
[25] Möller B, Löfstrand B, Karlsson M. "An overview of the HLA evolved modular FOMs". *Spring Simulation Interoperability Workshop. Norfolk, Virginia: Simulation Interoperability Standards Organization,* 2007.
[26] Möller B, Gustavson P, Lutz B, Löfstrand B. "Making your BOMs and FOM modules play together". *Fall Simulation Interoperability Workshop.* 2007.
[27] Möller B, Dahlin C. "A first look at the HLA evolved Web service API". *Euro Simulation Interoperability Workshop. Stockholm, Sweden: Simulation Interoperability Standards Organization,* 2006.
[28] Wang WG, Yu WG, Li Q, Wang WP, Liu XC. "Service-oriented high level architecture". *Euro Simulation Interoperability Workshop. Simulation Interoperability Standards Organization,* 2008.
[29] Moradi F, Ayani R, Mokarizadeh S, Shahmirzadi GHA, Tan G. "A rule-based approach to syntactic and semantic composition of BOMs". *Distributed Simulation and Real-Time Applications, 2007. DS-RT 2007. 11th IEEE International Symposium.* 2007. 145-155.
[30] Mojtahed V, Andersson B, Kabilan V, Zdravkovic J. "BOM++, a semantically enriched BOM". *Spring Simulation Interoperability Workshop. Simulation Interoperability Standards Organization,* 2008.
[31] Chase T. "Real-time platform reference (RPR) base object model (BOM) composability standard for enabling interoperability". 2007/2008-08-08. http://www.boms.info/Documents/RPR_BOMs/RPR_BOMs.ppt.
[32] Gustavson P, Chase T, Wilson M. "Building composable bridges between the conceptual space and the implementation space". *Winter Simulation Conference.* 2008.
[33] Tolk A, Muguira JA. "M&S within the model driven architecture". *The Interservice/Industry Training, Simulation & Education Conference (I/ITSEC). Orlando, Florida: Defense Acquisition University Press,* 2004.
[34] Fan C. "DDSOS: A dynamic distributed service-oriented modeling and simulation framework" Thesis. *United States -- Arizona: Arizona State University*, 2006.
[35] Zeigler BP, Hammonds PE. "Modeling & simulation-based data engineering: Introducing pragmatics into ontologies for net-centric information exchange". *New York, USA: Academic Press*, 2007.
[36] Mittal S. "DEVS unified process for integrated development and testing of service oriented architectures" Thesis. *United States -- Arizona: The University of Arizona*, 2007.
[37] Möller B, Morse KL, Lightner M, Little R, Lutz B. "HLA evolved – a summary of major technical improvements". *Fall Simulation Interoperability Workshop. Simulation Interoperability Standards Organization,* 2008.

**BIOGRAPHIES**

**WENGUANG WANG** is a Ph.D. candidate/student at College of Information Systems and Management, National University of Defense Technology (NUDT), China. He is a member of SISO and CASS (Chinese Association for System Simulation). His research interests are service-oriented simulation, High Level Architecture, DEVS, simulation composability and interoperability etc. His e-mail address is <wgwangnudt@gmail.com>.

**ANDREAS TOLK** is an Associate Professor for Engineering Management and Systems Engineering at Old Dominion University (ODU). He is also a Senior Research Scientist at the Virginia Modeling Analysis and Simulation Center (VMASC). He holds a M.S. in Computer Science (1988) and a Ph.D. in Computer Science and Applied Operations Research (1995). He is a member of SCS and SISO. His e-mail address is <atolk@odu.edu>.

**WEIPING WANG** is a Professor at National University of Defense Technology (NUDT), China. He is the founder of Systems Simulation Lab, College of Information Systems and Management, NUDT. He has over twenty years of experience in systems modeling and simulation community. He is a Ph.D supervisor. His research interests are systems simulation, system of systems engineering, simulation based acquisition, simulation composability and interoperability etc. His email address is <wangwp@nudt.edu.cn>.